\documentstyle {laa}
\input {psfig}

\def\rhomean{{\bar\rho}}
\def\Ho{H_o}
\def\inu{i_\nu}
\def\jnu{j_\nu}
\def\Snu{S_\nu}
\def\M15{M_{15}}
\def\T15{T_{15}}
\def\ymean{<y>}

\def\Mdot{M_\odot}
\def\nbare{{\bar n}_e}

\title{The Sunyaev-Zel'dovich Effect and the Value of $\Omega_o$}
\author{Domingos Barbosa\inst{1,2}, James G. Bartlett\inst{1}, Alain 
Blanchard\inst{1}, Jamila Oukbir\inst{3}}
\institute{$^1$Observatoire astronomique de Strasbourg, U.L.P., 
11, rue de l'Universit\'e, F67 000 Strasbourg, FRANCE\\  
Unit\'e associ\'ee au CNRS\\
$^2$Centro de Astrof\'{\i}sica da Universidade do Porto, 
Rua do Campo Alegre 823, 4150 Porto, PORTUGAL\\
$^3$ SAp, CE-Saclay, 91191 Gif-sur-Yvette Cedex, FRANCE}
\begin{document}
\maketitle

\begin {abstract}
We consider the Sunyaev-Zel'dovich (SZ) effect as a probe of $\Omega_o$:
Using a self-consistent modeling of X-ray clusters, 
we examine the dependence of both the mean Compton $y$ parameter 
and the SZ source counts on $\Omega_o$.  These quantities
increase with decreasing $\Omega_o$ due to the earlier epoch of structure 
formation in low--density cosmogonies; the results depend only 
on the quantity of gas heated 
to the virial temperature of collapsed objects and are independent 
of the spatial distribution of the gas in the potential wells.
Specifically, we compare two models which reproduce the 
present-day abundance of clusters - a biased, critical universe and
an unbiased, open model with $\Omega_o=0.3$.  We find that the mean
$y$ parameter approaches the current FIRAS limit of 
$y<2.5\times 10^{-5}$ for the open model, demonstrating
the importance of improving this limit on spectral distortions, and
that the SZ source counts and corresponding
redshift distribution differ significantly between the two 
cosmogonies;  millimeter surveys covering a large
area should thus provide interesting constraints on the density 
parameter of the Universe and on the evolution of the heated gas fraction
in virialized objects.

\end {abstract}

\section {Introduction}

	One of the fundamental differences between high-- and low--density 
cosmogonies concerns the rate of structure formation.  If we 
measure the density of the Universe by the dimensionless 
parameter $\Omega_o\equiv 8\pi G \rhomean /3\Ho^2$, where $\Ho$ 
is the current value of the Hubble constant, then 
the linear growth factor, $D_g(z)$, which governs the
rate of structure formation, tends to a constant for redshifts
$1+z<1/\Omega_o$.  In other words, the objects and structure
observed today exist relatively unchanged out to redshifts of 
order $1/\Omega_o-1$. This has been illustrated and proven in a formal 
way by Oukbir and Blanchard (1992,1995):  The redshift distribution 
of clusters of a given mass is almost
independent of the primordial power spectrum, but depends 
sensitively on the cosmic density, with a tail extending to large
redshifts for low $\Omega_o$.  This provides us with a probe
of $\Omega_o$.

	Because of their rarity, which is naturally interpreted 
in gaussian theories of structure formation as the result
of their being density perturbations on the tail of
the probability distribution, galaxy clusters are particularly 
sensitive to differences in the linear growth rate.  The 
discovery of significant numbers of clusters at redshifts greater
than 1 would be a strong argument for a low--density universe.
The problem is finding, if they exist, such high redshift
clusters.  Optical identifications of clusters become quite 
difficult at large redshifts due to the large number of projected 
foreground and background galaxies at the same magnitudes as the
cluster galaxies; moreover, our lack of understanding 
of galaxy evolution and its possible dependence on environment 
raise doubts on the interpretation, in the present context,
of optically detected, high redshift clusters.
Identifications based on the X-ray emission 
of the intra\-cluster medium (ICM), heated to the virial temperature 
of the gravitational potential well, fare better in this regard, 
but the X-ray luminosity, and thus the detectability of 
a cluster, strongly depends on the core radius of the
gas distribution; the $n^2$ dependence of the free-free 
emissivity guarantees that the core region of the cluster
dominates the total luminosity.  The problem is that the physics
determining this core radius is not well understood.

	The same hot gas responsible for the X-ray emission
also produces a spectral distortion of the cosmic microwave
background (CMB) known as the Sunyaev-Zel'dovich (SZ) effect
(Sunyaev \& Zel'dovich 1972).  Inverse
Compton scattering of the photons off the hot electrons in the
ICM causes the diffusion of low energy photons from the 
Rayleigh-Jeans region into the Wien tail of the (originally 
unperturbed blackbody) spectrum of the CMB.  This results
in a truly unique spectral signature -- at wavelengths longward
of 1.34 mm, the cluster appears as a decrement in the mean
sky brightness, while at shorter wavelengths one observes 
the cluster as a source of emission in excess of the mean
sky brightness.  Quantitatively, one expresses the change
in sky brightness relative to the mean CMB intensity, which we will 
refer to as the surface brightness of the source, although it
is negative at low frequencies, as
\begin {equation}
\label {bright}
\inu = y(\vec{\theta}) \jnu(x),
\end {equation}
where $\jnu$ describes the frequency dependence,
and the {\it Compton $y$ parameter}, which determines the magnitude 
of the distortion, is given by an integral along the line--of--sight 
through the cluster:
\begin {equation}
\label {comptony}
y \equiv \int dl \frac{kT}{m_ec^2} n_e \sigma_T.
\end {equation}
In this latter expression, $T$ is the temperature of the ICM 
(strictly speaking, of the electrons), $m_e$ is the electron
rest mass, $n_e$ is the electron density and $\sigma_T=6.65\times
10^{-25}\, $cm$^2$ is the Thompson cross section.  
Note that $y$ is a function of angular position $\vec{\theta}$
on the cluster image.  The spectral function $\jnu$
may be written in terms of the dimensionless frequency 
$x\equiv h_p\nu/kT_o$, where $h_p$ is Planck's constant and 
$T_o$ is the {\it current} temperature of the CMB -- 2.726K 
(Mather et al. 1994) -- as
\begin {equation}
\label {jnu}
\jnu(x) = 2\frac{(kT_o)^3}{(h_pc)^2} \frac{x^4e^x}{(e^x-1)^2}
	[\frac{x}{{\mbox{\rm tanh}}(x/2)} - 4].
\end {equation}

	The {\it flux} of a cluster, which is measured
in mJy ($=10^{-26}\,$ergs/s/cm$^2$/Hz), is the integral of the 
surface brightness $\inu(\vec{\theta})$ over the solid angle 
subtended by the cluster:
\begin {equation}
\label {fluxdens}
\Snu(x,M,z) = \jnu(x) D_a^{-2}(z) \int dV \frac{kT(M,z)}{m_ec^2} 
	n_e(M,z) \sigma_T;
\end {equation}
the integral is over the cluster volume and the angular distance 
$D_a(z)=2c\Ho^{-1}[\Omega_o z + (\Omega_o-2)(\sqrt{1+\Omega_o z} -1)]
/\Omega_o^2(1+z)^2$.  We see from this expression that the integrated
effect of a cluster, {\it i.e.} its flux, depends {\it only on the quantity 
of gas} at temperature $T$, which we will take to be the virial 
temperature, and {\it not on the spatial distribution of the gas}; 
as we have seen, this is in contrast to the X-ray flux.

	The other property of importance in the
present context is the distance independence of $y$: for a given
set of cluster properties, the value of $y$ remains independent
of redshift.  This implies that the mean distortion, averaged over
the full sky, equally weights the contributions 
from clusters at low and high redshifts.  Because 
the cluster population extends back to larger redshifts
in an open universe, we expect the mean $y$ to be
bigger than for a critical universe (Cavaliere et al. 1991;
Markevitch et al. 1991).  This is 
the characteristic which makes the SZ effect a potent probe of 
$\Omega$.  

	The situation is similar for the source counts: The fact that
one can observe clusters at high $z$ means that the counts
will be larger for a low--density universe (Korolev et al. 1986;
Markevitch et al. 1994).  As an instructive exercise, we may compare the
behavior of the SZ and X-ray fluxes with simple scaling laws: 
$f_S \sim MT/D_a^2(z)$; while for the X-ray flux we have 
$f_X \sim M n T^{1/2}/D_l^2(z)$ ($D_l$ is the luminosity distance).
We have written the baryonic mass of the cluster as $M$, its temperature
as $T$ and its gas density as $n$.  The dependence of the X-ray 
flux on gas density, and therefore on the spatial distribution
of the ICM, is manifest by the explicit appearence of $n$ in the
expression for $f_X$.
Using $T\sim M^{2/3} (1+z)$ (see below), we find that $f_S/f_X 
\sim (1+z)^4 T^{1/2}/n \sim (1+z)^{1.5}$, the last equality 
following for self-similar scaling, $n\propto (1+z)^3$.
In comparison with X-ray observations, one probes deeper in redshift 
with the SZ effect and, at the same time, avoids a sensitivity 
to the core radius of the gas distribution.

\section {Mean Spectral Distortion and Source Counts}

	For our calculations we adopt the Press-Schechter (1974) mass
function:
\begin {equation}
\label {ps}
\frac{dn}{d\ln M} d\ln M = \sqrt{\frac{2}{\pi}}\frac{\rhomean}{M}\nu(M,z)
	\bigg(-\frac{d\ln\sigma}{d\ln M}\bigg) e^{-\nu^2/2} d\ln M,
\end {equation}
in which $\rhomean$ is the comoving, or present--day, 
mass density of the Universe.
The quantity $\nu = \delta_c D_g(z)/\sigma(M)$ 
gives the height of the over-dense regions collapsing at redshift
$z$, with a {\it linear} density contrast of $\delta_c$, relative 
to the amplitude of the density perturbations at that epoch, 
$\sigma(M)/D_g(z)$, where $\sigma(M)$ is the amplitude of the density 
perturbations today.  The linear over-density corresponding to 
a just virialized object, $\delta_c$, depends weakly on redshift
if $\Omega_o \ne 1$ and equals $1.68$ if $\Omega_o=1$ (Oukbir \& Blanchard
1995).

	This formula demonstrates quantitatively the origin
of the particular sensitivity of clusters to the linear growth rate.
Clusters typically represent perturbations of several $\nu$ today,
and thus rapidly disappear as the average perturbation amplitude,
$\sigma(M)/D_g(z)$, decreases towards higher $z$.

\begin {figure*}
\psfig {figure=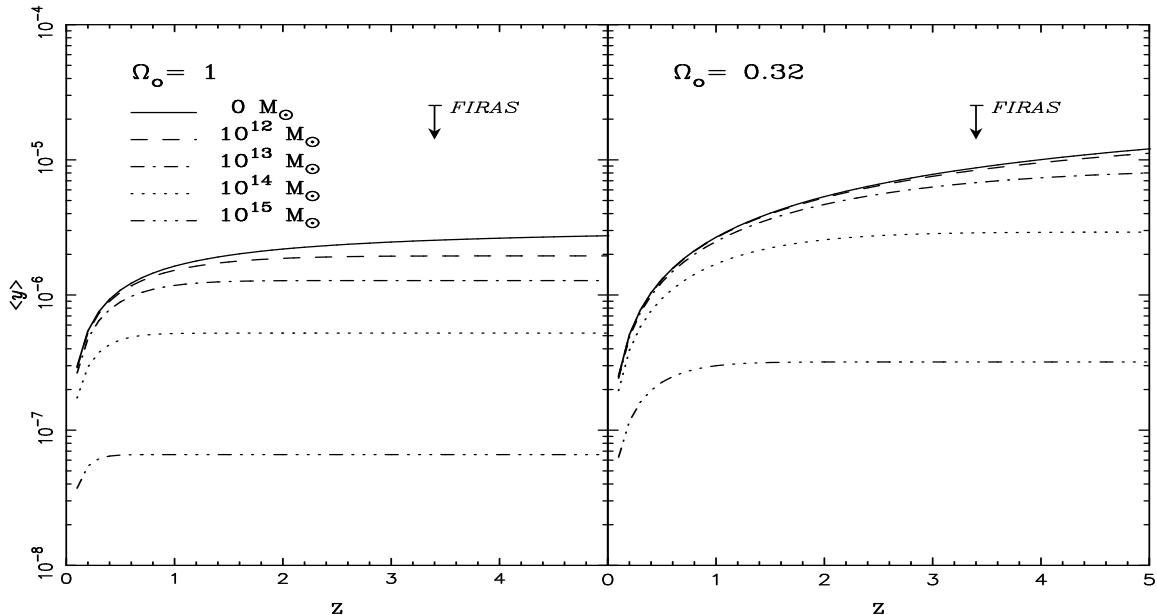,width=18cm,height=9cm,angle=270}
\caption {The mean Compton $y$ parameter for the two models; in 
both cases we have assumed $h=0.5$.  The different curves for each
$\Omega_o$ correspond to different low mass cutoffs, as identified
in the legend.  Compton cooling becomes important beyond a redshift 
of about 5; for this reason we restrict consideration
to the redshifts shown.}
\end {figure*}

	As per the discussion in the Introduction, the mean
Compton distortion depends only on the temperature distribution
of the baryons, not on their spatial distribution:
$$
\ymean = \int \sigma_T cdt \int dT \frac{d\nbare}{dT} 
		\frac{kT}{m_e c^2} 
$$
\begin {equation}
\label {meany} 
= f_B\int dz \frac{dt}{dz} c\sigma_T (1+z)^3 \int d\ln M \frac{dn}{d\ln M}
	\frac{\chi M}{m_p} \frac{kT(M,z)}{m_ec^2},
\end {equation}
where, in the first line, $\nbare$ is the mean electron 
density.  In the second line, $m_p$ is the proton mass, 
$\chi$ is the number of electrons per baryon, and the extra 
factor of $(1+z)^3$ changes 
the comoving density in eq. \ref{ps} to a proper density.  
For a primordial gas composition, $\chi=0.88$.

	Detailed X-ray imaging of clusters, such as Coma, indicate baryon 
fractions $f_B$ as large as $\sim 0.05h^{-3/2}$ ($h\equiv 
\Ho/100\,$km/s/Mpc)  (Briel etal. 1992; Mushotsky 1993).  This poses 
a problem for flat cosmologies (White et al. 1993a) if one accepts 
the baryon density dictated by primordial nucleosynthesis -- 
$\Omega_Bh^2 = 0.012$ (Walker et al. 1991) -- for one expects 
$f_B=\Omega_b/\Omega_o$.  In the open model we have chosen, 
with $\Omega_o=0.3$ and $h=0.5$, presented below, the 
observed baryon fraction of clusters is consistant with
the theory of primordial nucleosynthesis.  
To isolate the effects of $\Omega_o$, 
we choose the {\it same} value of $f_B=0.2$ for the critical
model, even though this represents a violation of primoridal
nucleosynthesis theory.  
One may easily scale the results presented in our figures
to any desired baryon fraction by multiplying $\ymean$
by $f_B/0.20$ and by sliding the counts at a 
given $S_\nu$ to $(f_B/0.20)S_\nu$, everything else 
being equal. 

	Straightforward scaling arguments tell us that the temperature 
of a virialized object $T\sim M/R$, and that the virial
radius $R\sim M^{1/3}\Delta^{-1/3}\Omega_o^{-1/3}(1+z)^{-1}$.
The spherical collapse model permits the calculation of
the (non-linear) over-density, $\Delta(z)$, at the time of virialization and
of the other numbers needed to find the coefficients of 
these scaling relations.  The result for a critical universe 
is in good agreement with hydrodynamical simulations (Evrard 1990), 
which give $T(M,z) = \T15 \M15^{2/3} (1+z)$ with $\T15=6.4h^{2/3}$ 
keV; it is convenient to express the mass $M$ of a cluster in terms
of $\M15\equiv M/10^{15}\Mdot$.  Taking this as a normalization, 
we henceforth adopt a temperature--mass relation of the 
form: $T(M,z) = \T15(z) \M15^{2/3} (1+z)$ with $\T15(z) = 
(6.4\, \mbox{\rm keV}) h^{2/3} \Omega_o^{1/3} (\Delta(z)/178)^{1/3}$ 
($\Delta=178$ in a critical universe, independent of $z$).
 
	To perform the integral in eq. \ref{meany}, we must also choose
the form and normalization of the power spectrum, $\sigma(M)$, at the 
present epoch.  In this {\it paper}, we use a power--law form
for the power spectrum - $\sigma(M)=M^{-\alpha}$ - which translates
into $P(k)\propto k^n$, with $\alpha=(n+3)/6$, in Fourier space.  
One usually expresses the normalization of
the power spectrum in terms of the amplitude of the density
perturbations in spheres of radius $8h^{-1}\,$ Mpc, or $\sigma_8$.
Following Oukbir \& Blanchard (1995), we determine $n$ and 
$\sigma_8$ by transforming the PS mass function with the 
temperature--mass relation and fitting to the observed
temperature distribution at the present epoch as determined
by Edge et al. (1990) and Henry \& Arnaud (1991).  For a
critical universe with $h=0.5$, we find $n=-1.85$ and $\sigma_8=0.6$
(Henry \& Arnaud 1991; Blanchard \& Silk 1992; Bartlett \& Silk
1993; White et al. 1993b; Oukbir \& Blanchard 1995).
Note that a cold dark matter model with $h=0.5$ has $n=-1$ on
cluster scales and thus does not provide a good fit to the
shape of the temperature distribution function (in addition,
when the model is normalized to the COBE fluctuation 
amplitude, $\sigma_8$ is $\sim 1$, {\it i.e.} much too 
large for $\Omega_o=1$).

	Because the mean density determines the mass contained in 
spheres of radius $8h^{-1}\,$Mpc, $\sigma_8$ will depend on 
$\Omega_o$ if one is trying to fit a given cluster temperature 
distribution.  For concreteness, we choose an unbiased 
low--density model; fixing $\sigma_8=1$ while adjusting 
$n$ and $\Omega_o$, Oukbir \& Blanchard (1995) find that $n=-1.42$ 
and $\Omega_o=0.32$ provide the best fit to the temperature 
distribution function.  Here again we assume $h=0.5$.
This will serve as our example of a low--density universe.

\begin {figure*}
\psfig {figure=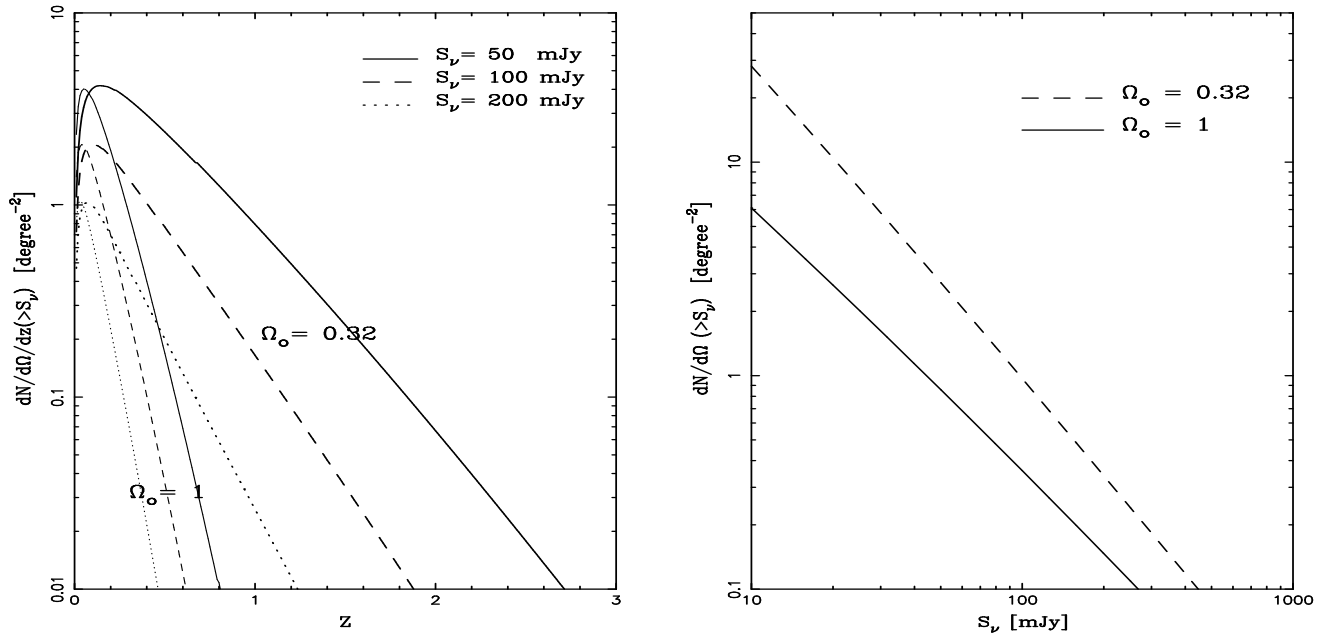,width=18cm,height=9cm,angle=270}
\caption {The SZ source counts (right) and redshift distribution for
several values of $S_\nu$ (left).  The calculation is for a
wavelength of 0.75 mm and an $h=0.5$.}
\end {figure*}

	We prefer this fitting to the temperature function
over one using the X-ray luminosity function because the 
temperature--mass relation is essentially based on energetics,
while the core radius strongly influences the X-ray 
luminosity of a cluster.
Since each model is now normalized to the present-day 
cluster abundance, a comparison of their predictions
permits us to directly examine the influence of $\Omega_o$ 
on SZ observations.

	We compare the mean CMB spectral distortion in the two
models in figure 1, where we show the mean $y$ value integrated out to 
redshift $z$ as a function of $z$.  The various curves for each 
value of $\Omega_o$ correspond to different lower bounds on the mass 
integral in eq. \ref{meany}. With the cutoff mass set to zero, 
the integrated $\ymean$ in an 
Einstein--de--Sitter universe reaches a value of $\sim 3\times 10^{-6}$, 
while in the low--density universe it reaches an asymptotic
value at large redshift of more than $5\times 10^{-5}$! - well 
above the present FIRAS limit of $y<2.5\times 10^{-5}$ 
(Mather et al. 1994).  
However, one should take into account that the gas 
in low temperature halos cools within a Hubble time, 
thereby providing a low mass cutoff to the integral; 
this mass is typically of the order of $10^{12}$ M$_{\odot}$, 
almost independent of redshift (Blanchard et al. 1992).
One should also take into account Compton cooling, which becomes 
efficient at redshifts greater than $\sim 5$.
Nevertheless, in a low-density universe at redshift 5, the 
difference between a cutoff mass of $10^{12}\Mdot$, or even 
one as large as $10^{13}$, and a zero mass cutoff is quite small;
and the predicted $\ymean$ reaches $10^{-5}$ - just below 
the current limit.  We conclude that the expected mean Compton distortion 
for a low--density Universe remains just below the current 
limits from FIRAS and that an effort to reduce the limits on
$\ymean$ would provide an interesting constraint on low
density models (or a detection!).  In a critical universe, 
on the other hand, $\ymean$ should be about one order of magnitude
below the present limit; with an $f_B$ as predicted by primordial
nucleosynthesis, this becomes even lower, by a factor of $\sim 3$.

	Next consider the number of SZ sources 
on the sky brighter than a given threshold $S_\nu$ (expressed
in mJy).  We simply integrate the PS mass function over all 
objects with a flux density brighter than $S_\nu$:
\begin {equation}
\label {countsint}
\frac{dN}{d\Omega}(>S_\nu) = \int dz\frac{dV}{dzd\Omega}
	\int_{M_{min}(S_\nu,z)} dM \frac{dn}{dM}.
\end {equation}  
In this equation, $M_{min}$ is the
mass corresponding to the threshold, $S_\nu$, as determined
by relation \ref{fluxdens}:
\begin {equation}
\label{fluxdens2}
S_{\nu} = (8\,\mbox{\rm mJy}\, h^{8/3}) f_\nu(x) f_B 
	\Omega_o^{1/3}M_{15}^{5/3}[\frac{\Delta(z)}{178}]^{1/3}
	(1+z) D^{-2}(z), 
\end {equation}
where $f_\nu(x)$ and $D(z)$ are the dimensionless parts of eq. 
\ref{jnu} and the angular distance, respectively (in both cases,
without the factor 2).
This procedure gives the number of objects per solid angle on the
sky as a function of the {\it total} flux density of the
objects - in other words, it assumes that the sources are
unresolved in the experimental beam.

	We present the calculated source counts at 0.75 mm  
for the two cosmogonies in figure 2.  The redshift 
distribution for each model is also shown, in the adjacent panel.  
In a low--density universe one predicts a larger number of 
sources at a given threshold and a broader distribution in
redshift, extending out to higher redshifts, than would be the case
in a critical universe.  Thus, we see the effect of the 
geometry on the linear growth rate: in a low--density universe,
clusters exist out to larger redshifts and, hence, appear in
greater numbers in the source counts.
In a manner similar to X-ray selected clusters, 
the redshift distribution of SZ clusters
offers a unique way to determine the mean density of the universe -
the {\it shape} of this distribution provides a robust 
indicator of the mean density.  The advantage of SZ cluster
catalogs, over X-ray selected samples, is that the 
detection criteria are insensitive to the gas core radius 
and its evolution.

\section {Discussion}

	The Sunyaev-Zel'dovich effect is a natural complement to 
optical and X-ray studies of galaxy clusters, and, in contrast,
suffers neither from projection effects nor from a sensitivity
to the spatial distribution of the hot, intracluster gas.  It simply 
probes the quantity of gas heated to the virial temperature of collapsed
objects (we consider here only unresolved sources).  In this 
light, the SZ effect serves as the more robust probe of the evolution
of this gas fraction and of the value of $\Omega_o$, the latter
determining the epoch of structure formation: In a low--density
universe, structure must exist back to a redshift $1+z \sim 1/\Omega_o$,
for the linear growth of fluctuations stops at lower redshifts due
to the more rapid expansion of space.  A low--density universe
thus provides a longer baseline over which to integrate the 
cluster SZ effect and should produce both a larger $\ymean$ and
greater number counts, with a larger mean redshift, than a critical 
universe.  This is what we see in figures 1 and 2.

	As demonstrated in figure 1, the mean spectral distortion of 
the CMB depends surprisingly little on the lowest masses one includes 
in the calculation, at least if one agrees that masses as low as,
say, $10^{13}\Mdot$ can safely be included in the computation. 
Then, the $\ymean$ contributed by structure back to a redshift 5 in
our low--density model reaches $\sim 10^{-5}$; beyond a redshift
of $\sim 5$, Compton cooling dominates and should effectively eliminate
the hot gas in most structures.  This is only a factor of 
2--3 below the current FIRAS limit; lower cosmic densities
would predict distortions even closer to the FIRAS limit.
In comparison, a critical universe presents a very small 
spectral distortion, at most $few \times 10^{-6}$.  This 
dependence of $\ymean$ on $\Omega_o$ has already been
remarked on by Cavaliere et al (1991) and Markevitch
et al (1991).  Further effort to increase the sensitivity to Compton
distortions would provide useful constraints on low--density
models, or, perhaps, the more interesting possibility of a detection.
Motivation for pushing down limits on $\ymean$ also comes from
consideration of the physical state of a uniform intergalactic 
medium: One could probe the quantity of gas in the medium 
(Bartlett et al. 1995).

	As first noted by Korolev et al (1986), cluster source 
counts also prove to be a useful probe of $\Omega_o$.  The counts
expected in our two fiducial models are shown in figure 2.  
A SZ selected cluster catalog would be the 
best choice for studying the evolution of the gas fraction in 
clusters and $\Omega_o$.  Of particular importance is the difference
in the number of detected sources, at a given threshold, between low and 
high density cosmologies and the fact that the detected 
clusters will extend out to much larger redshifts in the low 
density case - a smoking gun for a low value of $\Omega_o$ (which is
to say that if such high--redshift clusters exist, this would
be a very strong indication for a low--density universe, although
if they do not exist, one might still argue that $\Omega_o$ is
low, but that the hot gas fraction drops rapidly with look-back
time).  A study of this kind requires mapping a significant 
portion of the sky at millimeter and sub-millimeter wavelengths.
This would be possible with a satellite similar to the SAMBA/COBRAS
mission proposed to the European Space Agency; such a mission
could cover most of the sky with a sensitivity to SZ point sources
on the order of $\sim 100\,$mJy.  From figure 2, we see that this is
a sensitivity perfectly matched for detailed studies of the
evolution of clusters and $\Omega_o$.

	We emphasize what we feel to be an important aspect of
the present work: that in comparing a high and a low--density
model, to distinguish the influence of $\Omega_o$, we have
insisted that the power spectra in each case reproduce the
present-day abundance of galaxy clusters as measured by
the temperature distribution function.

	We have not considered the fluctuations induced
in the CMB by the SZ effect in this paper, but leave
this to a future work (Barbosa et al. 1996).  This issue
has been addressed by numerous authors using a variety
of approaches (Shaeffer \& Silk 1988; Cole \& Kaiser 1989;
Bond \& Meyers 1991; Markevitch et al. 1992; Bartlett
\& Silk 1994; Colafrancesco \& Vittorio 1994).  In contrast
to the mean CMB spectral distortion and the SZ source 
counts, the induced temperature fluctuations depend 
on the spatial distribution of the gas within clusters
and on the distribution of the clusters themselves.  
 
	Finally, we remark that flat, low--density models, with 
a non-zero cosmological constant, would produce distortions and 
counts somewhere in between the predictions presented here - 
linear growth of density perturbations in such models turns off
at lower redshifts than in the equivalent open model, but still higher
than for the $\Omega_o=1$ case.  

\acknowledgements
We are happy to thank C. Lineweaver for comments and useful discussions.  
D.B is supported by the Praxis XXI CIENCIA-BD/2790/93 grant 
attributed by JNICT, Portugal.

\newcommand{\asb}[3]{#1, Ann. sco. sci. Bruxelles #2\rm, #3}
\newcommand{\nat}[3]{#1, Nature #2\rm, #3}
\newcommand{\np}[3]{#1, Nucl. Phys. #2\rm, #3}
\newcommand{\apss}[3]{#1, Ap\&SpS #2\rm, #3}
\newcommand{\apj}[3]{#1, ApJ #2\rm, #3}
\newcommand{\aj}[3]{#1, AJ #2\rm, #3}
\newcommand{\apjl}[3]{#1, ApJ #2\rm, L#3}
\newcommand{\apjsup}[3]{#1, ApJ Supp. #2\rm, #3}
\newcommand{\aeta}[3]{#1, A\&A #2\rm, #3}
\newcommand{\annrev}[3]{#1, ARA\&A #2\rm, #3}
\newcommand{\mnras}[3]{#1, MNRAS #2\rm, #3}
\newcommand{\jrasc}[3]{#1, JRASC #2\rm, #3}
\newcommand{\physlettb}[3]{#1, Phys. Lett. B #2\rm, #3}
\newcommand{\physrevlett}[3]{#1, Phys. Rev. Lett. #2\rm, #3}

\end {document}